\documentclass{article}

\usepackage[preprint,nonatbib]{neurips_2020}

\usepackage[utf8]{inputenc} 
\usepackage[T1]{fontenc}    
\usepackage{hyperref}       
\usepackage{url}            
\usepackage{booktabs}       
\usepackage{amsfonts}       
\usepackage{nicefrac}       
\usepackage{microtype}      

\usepackage{amssymb}
\usepackage{caption}
\usepackage{graphicx}
\graphicspath{ {./images/} }
\usepackage{subcaption}
\usepackage{wrapfig}

\title{Generalising sequence models for epigenome predictions with tissue and assay embeddings}

\author{%
  Jacob Deasy\thanks{Corresponding author, \texttt{jacob.t.deasy@gsk.com}.} \\
  GSK.ai
  \And
  Ron Schwessinger \\
  GSK.ai
  \And
  Ferran Gonzalez \\
  GSK.ai
  \And
  Stephen Young \\
  GSK.ai
  \And
  Kim Branson \\
  GSK.ai
}

\begin{document}

\maketitle

\begin{abstract}
  Sequence modelling approaches for epigenetic profile prediction have recently expanded in terms of sequence length, model size, and profile diversity. However, current models cannot infer on many experimentally feasible tissue and assay pairs due to poor usage of contextual information, limiting \textit{in silico} understanding of regulatory genomics. We demonstrate that strong correlation can be achieved across a large range of experimental conditions by integrating tissue and assay embeddings into a Contextualised Genomic Network (CGN). In contrast to previous approaches, we enhance long-range sequence embeddings with contextual information in the input space, rather than expanding the output space. We exhibit the efficacy of our approach across a broad set of epigenetic profiles and provide the first insights into the effect of genetic variants on epigenetic sequence model training. Our general approach to context integration exceeds state of the art in multiple settings while employing a more rigorous validation procedure.
\end{abstract}

\section{Introduction}\label{sec:Introduction}

Sequence models (SMs) for epigenetic state prediction have the potential to improve understanding of regulatory genomics while augmenting the role of \textit{in silico} experimentation. By mapping DNA sequences to targets computationally, epigenetic sequence models offer rapid mutagenesis across a range of experimental contexts \cite{basenji2,enformer}. When these models are constructed for interpolation, detailed epigenetic signals can be analysed in inaccessible tissues which are often missing or rare in existing databases \cite{epcot,ocelot}. Despite their recent success, DNA-based models for epigenetic feature prediction continue to add output tracks in a manner which limits scalability and bottlenecks performance. They also demonstrate poor correlation between individuals, indicating a lack of generalisation across sequence space \cite{huang2023personal,sasse2023far}.

Deep neural networks, comprising convolutional and Transformer-based modules \cite{lecun1995convolutional,vaswani2017attention}, currently provide state of the art (SOTA) correlation when predicting epigenetic signals from the reference genome \cite{epcot,enformer}. These models train across a large range of cell types and experimental contexts, where they recognise functionally-relevant regulatory motifs present in the DNA input sequence. However, such models share a latent representation of the sequence across all contexts because the increasing number of experimental settings considered is integrated into their output space. Although expanding the output features can lead to a broad understanding of functional genomics, it also imposes a representational bottleneck, ignores the fundamental diversity of cell states \cite{seq2cells}, and does not generalise to predicting unseen experimental settings. Therefore, integrating contextual information in a more flexible manner will lead to models with a truly scalable prediction space.

In this work, we present a new approach, the Contextualised Genomic Network (CGN), which:
\begin{enumerate}
  \item Exceeds state of the art cross-gene correlation for a broad spectrum of epigenetic contexts (tissue-assay pairs) while matching SOTA model size and context length.
  \item Employs a more rigorous validation scheme than prior approaches, generalising over held out individuals, tissues, and assays in a zero-shot manner.
  \item Is the first DNA to epigenome model using Epigenomes from four individuals (ENTEx) \cite{entex}, providing preliminary results on the impact of training epigenetic SMs with variants.
\end{enumerate}

\begin{figure}[h]
  \begin{subfigure}[b]{\textwidth}
    \centering
    \includegraphics[width=\textwidth]{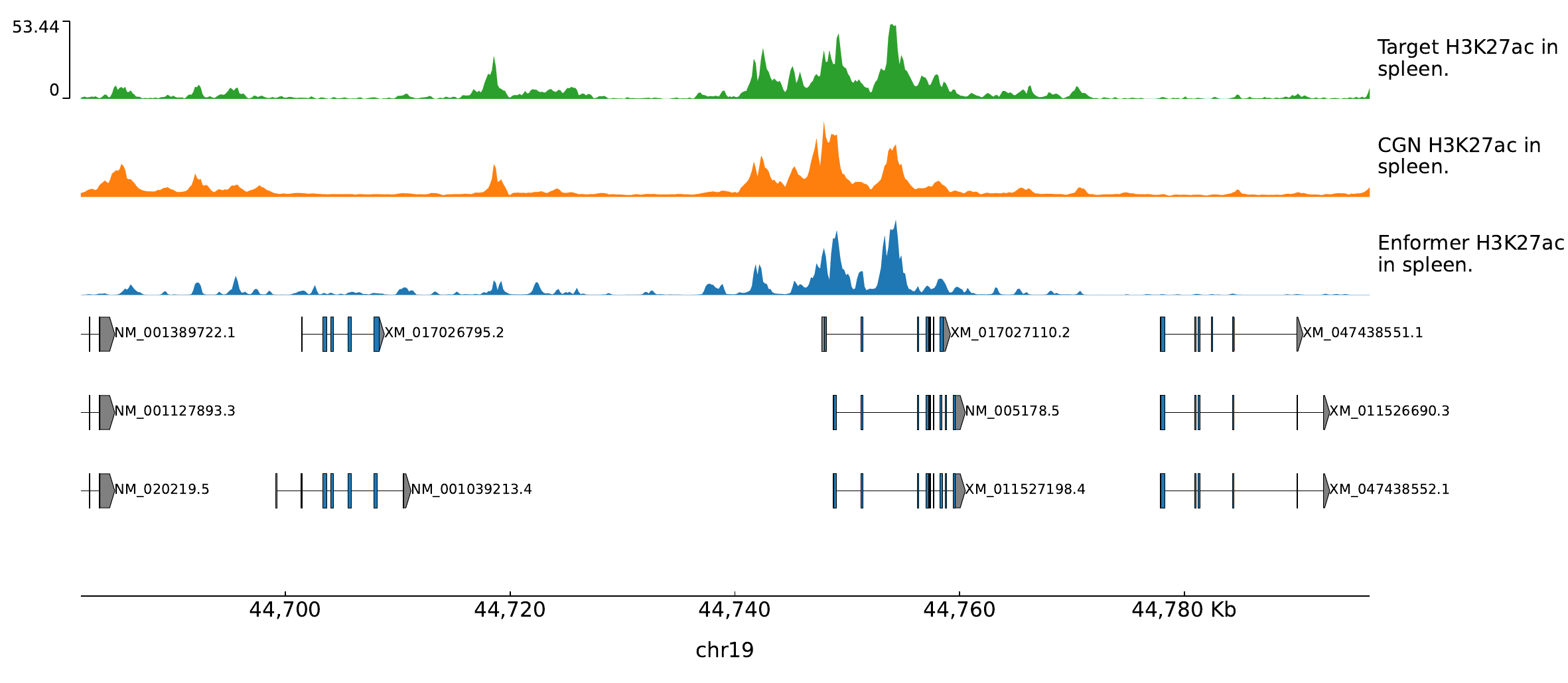}
    \caption{CGN's unitless prediction vs. Enformer's prediction for a tissue-assay pair that CGN has not seen\footnotemark[1].}
    \label{fig:...}
    \vspace{3mm}
  \end{subfigure}
  \subfloat[The CGN architecture.]{
    \includegraphics[clip,width=0.4\columnwidth]{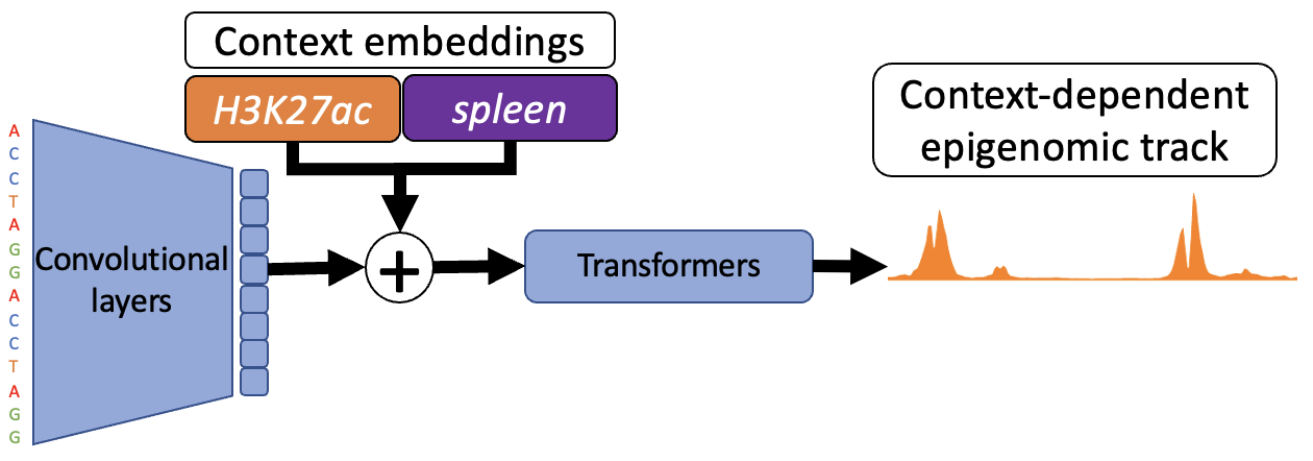}\label{fig:architecture}%
  }

  \vspace{-23mm}

  \subfloat[Per-context comparison.]{\includegraphics[width=0.4\textwidth]{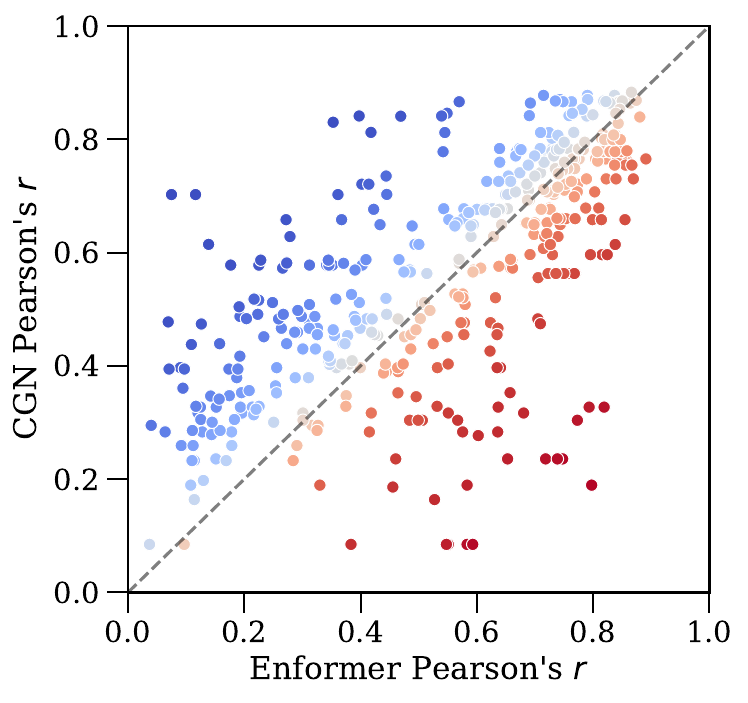}\label{fig:scatter_plot}}
  \hfill
  \begin{subfigure}[t]{0.6\textwidth}
    \centering
    \includegraphics[width=\textwidth]{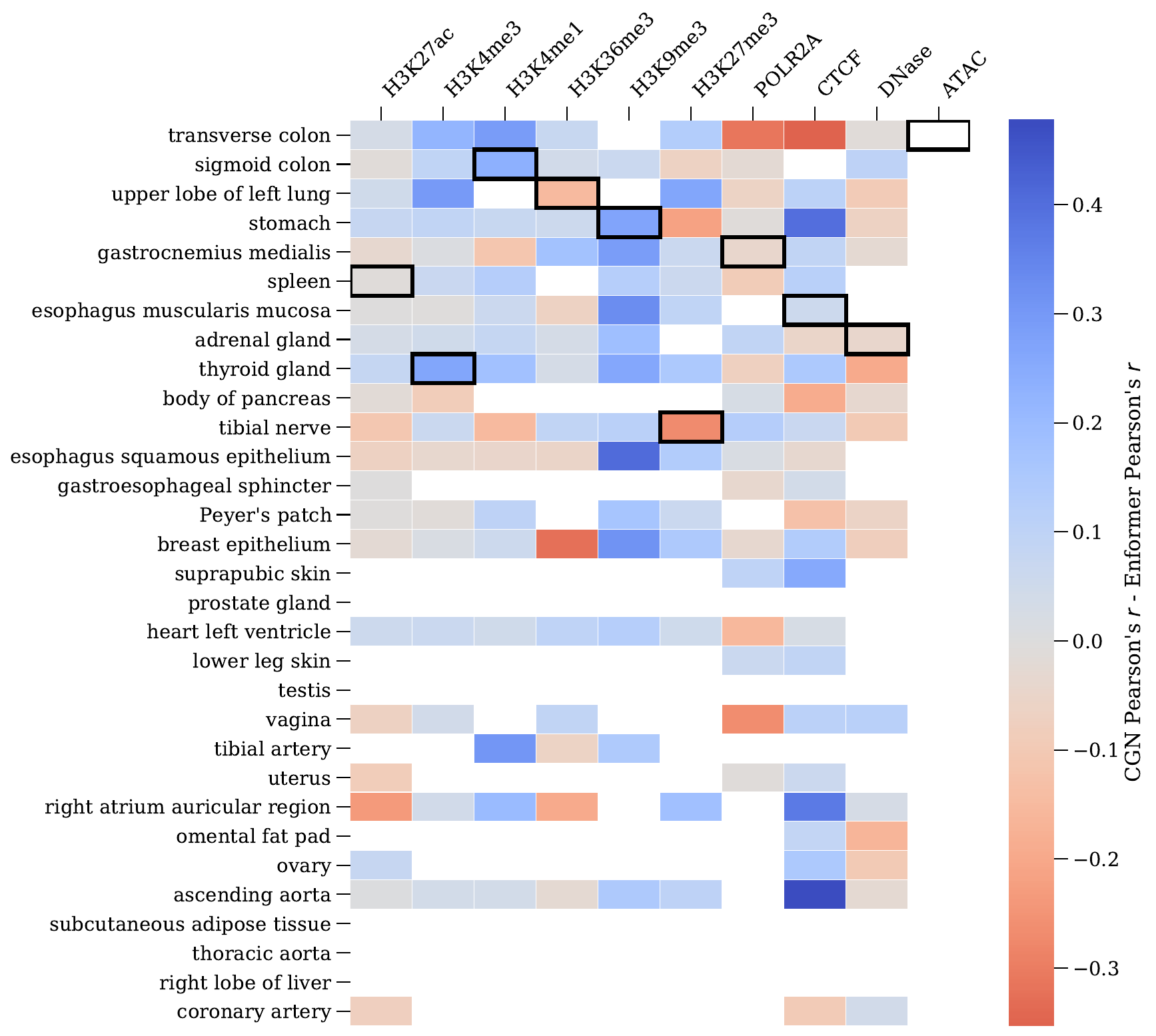}
    \caption{Improvements on comparable tissues \& assays.}
    \label{fig:heldout_donor_improvements}
  \end{subfigure}
  \caption{Test set CGN improvement over Enformer \cite{enformer} for contexts available for unseen ENTEx donor ENCDO271OUW. In (d), missing and black-border cells indicate missing and held out tissue-assay pairs, and Enformer performance is averaged over the tracks relevant to each pair.}
  \label{fig:heldout_donor}
\end{figure}

\footnotetext{Selected from the first 100 test samples as an example of regions where both models have high Pearson's $r$, Enformer $r=0.847$ and CGN $r=0.850$. Further examples are provided in the Appendix.}

\section{Results}
\label{sec:Results}

\subsection{CGN improves epigenetic feature prediction}
\label{sec:NGSNet improves epigenetic feature prediction}
We designed a new architecture, the Contextualised Genomic Network (CGN), to predict epigenetic features using context tokens to generalise to unseen tissue-assay pairs. Context tokens are common in natural language processing (NLP), in the form of prompts \cite{chainofthought}, as well as in multilingual models \cite{whisper}. Networks trained with contextual information learn, potentially simpler, conditional distributions which can be estimated by the entire network, rather than relying on a model head to delineate a shared latent representation \cite{multimodal}. For example, existing models try to predict both chromatin immunoprecipitation sequencing (ChIP-seq), Assay for Transposase-Accessible Chromatin using sequencing (ATAC-seq), and many other targets from a shared latent space with a low-capacity multi-layer perceptron head \cite{basenji2,enformer}. As these targets can derive from fundamentally different cellular processes and have drastically different distributions, this setup does not account for context-specific distal regulation, target normalisation, or model optimisation across diverse distributions. On the other hand, integrating experimental context in the input space allows powerful Transformer-based models to approximate epigenetic signals in a manner suited to the tissue and assay in question. By employing context tokens, we were able to improve model performance in the majority of cases while enabling generalisation to unseen tissue-assay pairs. This formulation bridges the gap between rigid SOTA models Basenji2 and Enformer and generalisable, but capacity-limited, interpolation models Avocado and Ocelot \cite{avocado,ocelot,eDICE}. We summarise comparable model properties of CGN in Table~\ref{tab:model_properties}.
\begin{table}[h]
  \caption{Input, output, and generalisation properties of existing models.}
  \label{tab:model_properties}
  \centering
  \small
  \begin{tabular}{ lrrcccc } 
    \toprule
    \textbf{Model} & \textbf{Input (bp)} & \textbf{Output (bp)} & \textbf{DNA only} & \textbf{Unseen context} & \textbf{Embeds context} & \textbf{Zero-shot} \\
    \midrule
    Basenji2 & 131,072 & 114,688 & \checkmark &            &            & \checkmark \\
    Enformer & 196,608 & 114,688 & \checkmark &            &            & \checkmark \\
    Epcot    & 1,600   & 1,000   &            & \checkmark & \checkmark &            \\
    Ocelot   & 300     & 25      &            & \checkmark &            & \checkmark \\
    CGN   & 196,608 & 114,688 & \checkmark & \checkmark & \checkmark & \checkmark \\
    \bottomrule
  \end{tabular}
\end{table}

\begin{figure}[h]
  \center
  \includegraphics[width=0.6\textwidth]{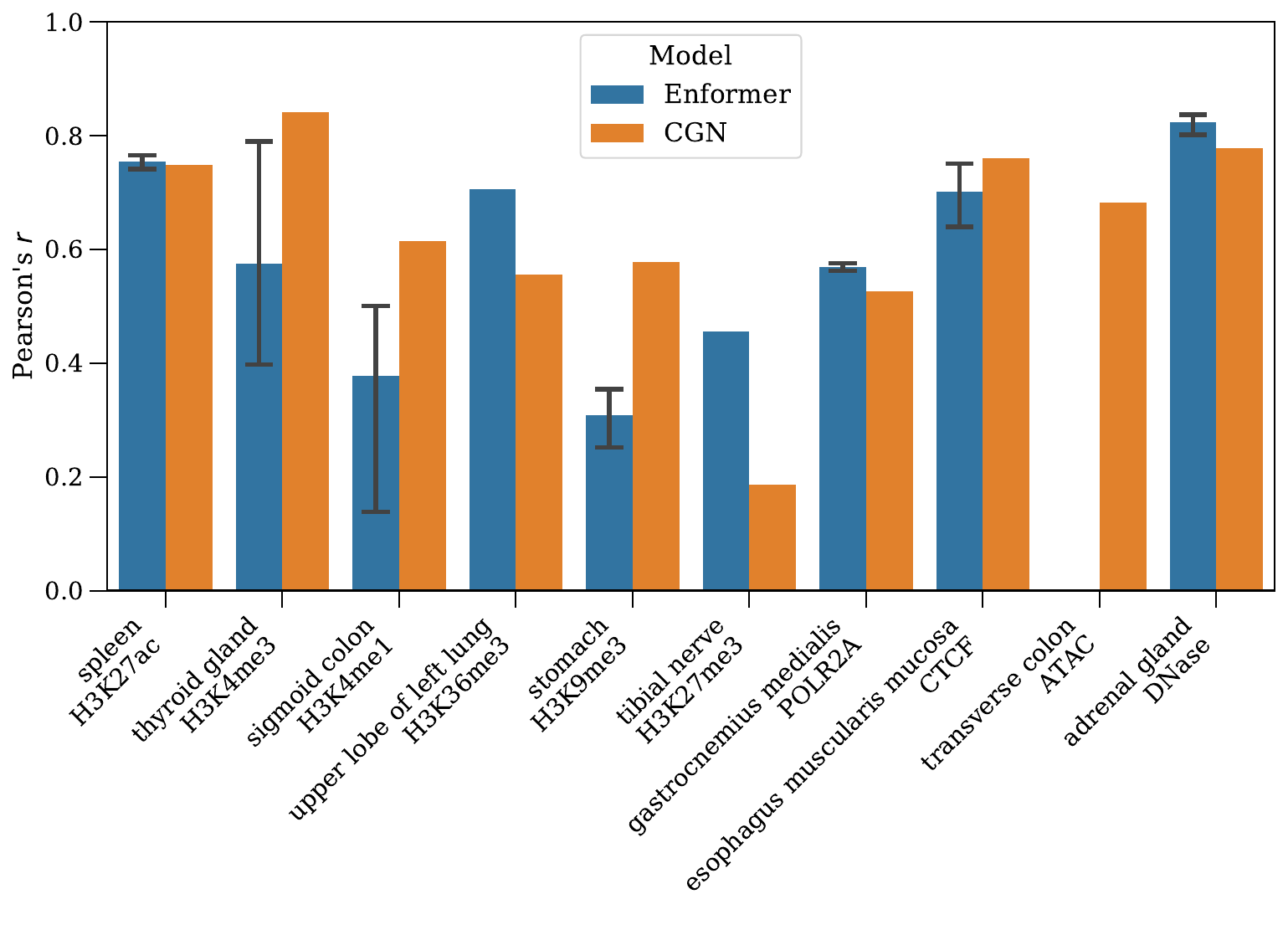}
  \caption{CGN and Enformer performance across contexts unseen by CGN. CGN is evaluated on unseen donor ENCDO271OUW from the ENTEx database. For contexts with multiple relevant Enformer outputs, Enformer performance is averaged across these tracks and a 95\% confidence interval around the mean is bootstrapped from 1,000 resamples.}
  \label{fig:heldout_pairs_model_comparison}
\end{figure}

\begin{figure}[h]
  \center
  \includegraphics[width=\textwidth]{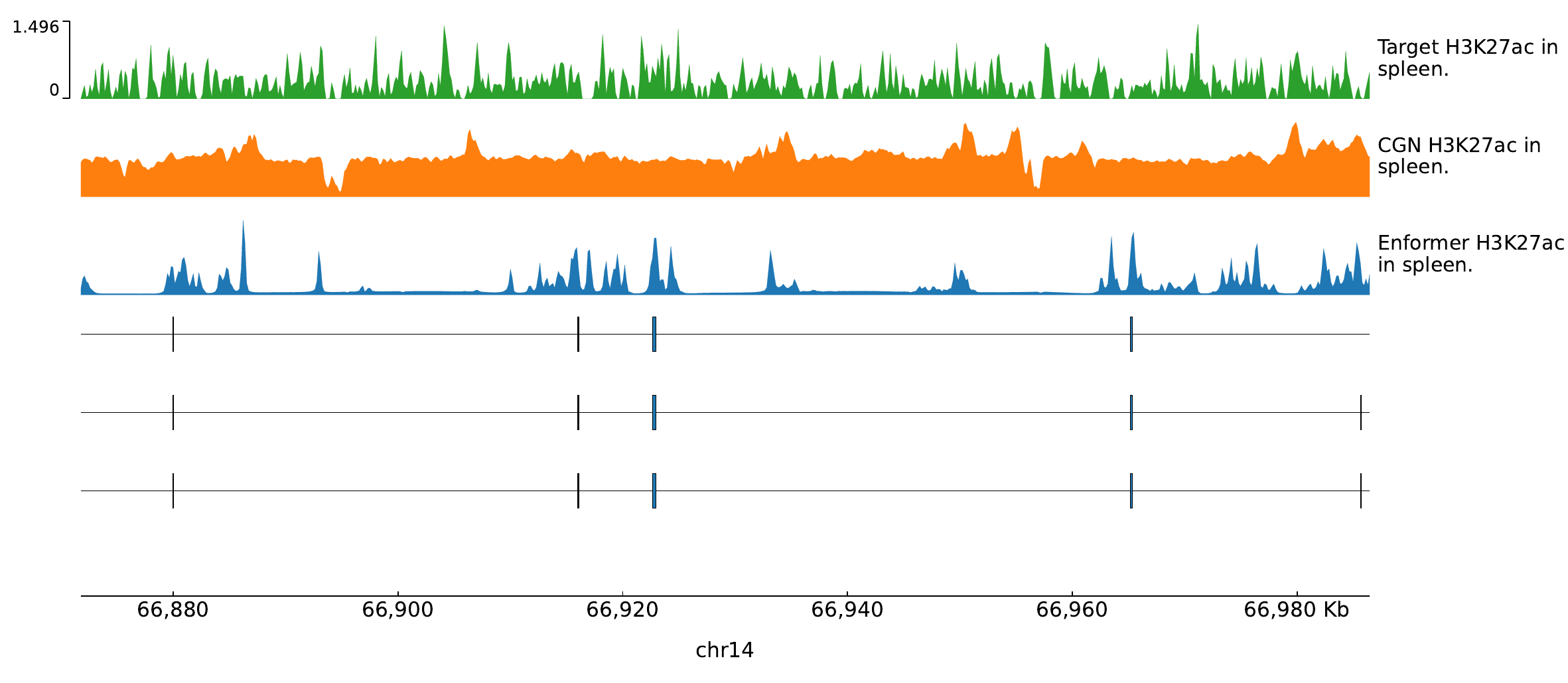}
  \caption{An example test set region where Enformer has low correlation (\textit{local} Pearson's $r=0.058$) and CGN avoids predicting spurious peaks (\textit{local} Pearson's $r=0.139$). We select this region (test set sample 6) as it is the first test set region where Enformer's local Pearson's $r$ is below 0.1 for this tissue-assay pair and because this trend exemplifies the trend across other samples.}
  \label{fig:bad_region}
\end{figure}

\begin{figure}[h]
  \center
  \includegraphics[width=\textwidth]{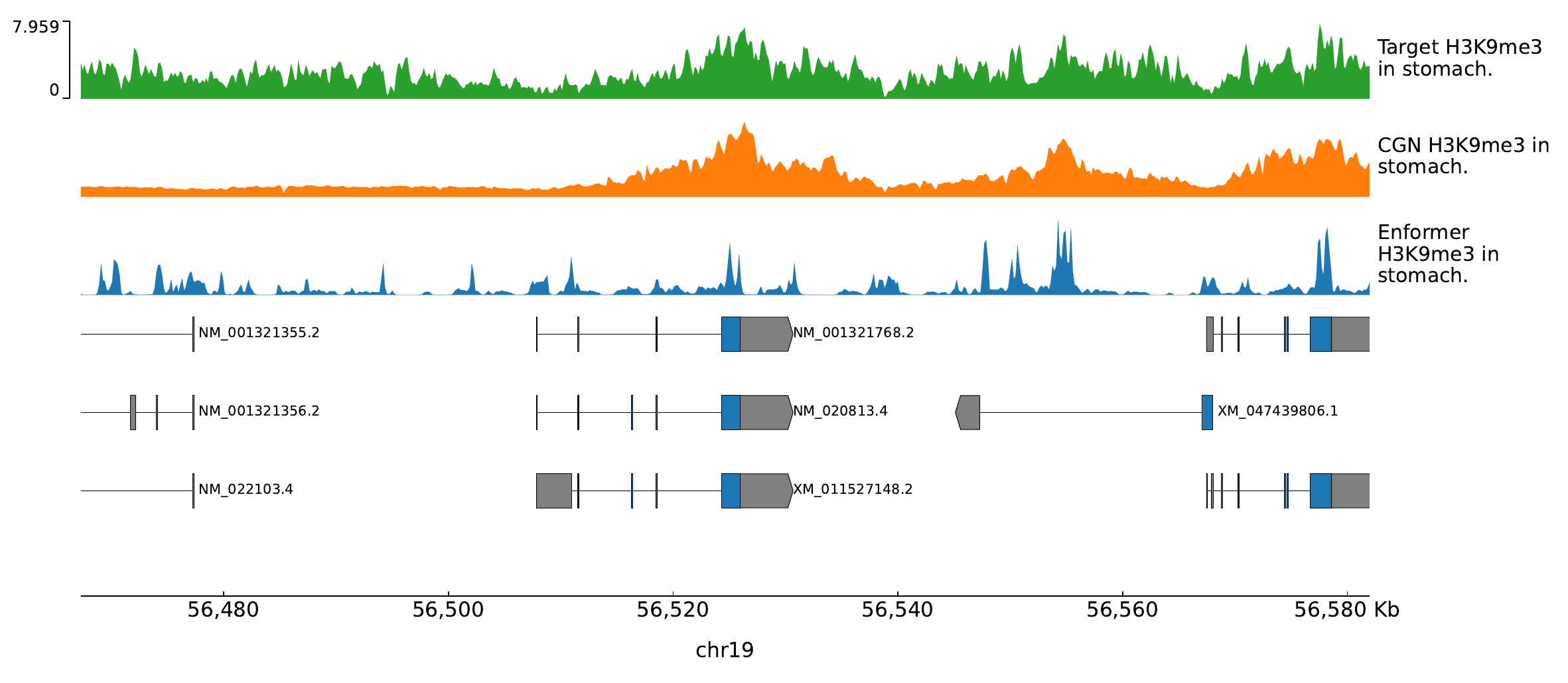}
  \caption{An example test set region where CGN more accurately captures broad repressive LFOC. Enformer local Pearson's $r=0.128$ and CGN local Pearson's $r=0.696$.}
  \label{fig:repressive}
\end{figure}

CGN outperforms the prior state of the art model Enformer across 98 out of 160 (61.25\%, Figure~\ref{fig:heldout_donor}) comparable contexts in terms of Pearson's $r$, with inference possible in a further 150 contexts, many of which are outside the scope of Enformer. This improvement includes 10 tissue and assay pairs which have never been seen by CGN. Additionally, the performance of CGN is measured in an unseen donor, representing the first inter-individual validation of epigenetic SMs, accounting for target distribution shifts and genetic variation. Performance improvements are spread throughout the tissues and assays present in ENTEx (Figure \ref{fig:heldout_donor_improvements}). In particular, Figure~\ref{fig:scatter_plot} highlights CGN's performance improvement on tracks where Enformer has low correlation, Figures~\ref{fig:bad_region} and \ref{fig:repressive} exemplify that this is mostly due to CGN's better avoidance of peak prediction for low-signal regions or targets. Performance improvements are predominantly correlated with the assay context, implying a dependency on per-assay target distributions, with the largest improvement seen in ChIP-seq experiments for histone mark H3K9me3 and architectural factor CTCF. We highlight performance in the randomly held out tissue-assay pairs in Figure~\ref{fig:heldout_pairs_model_comparison}. Despite never having seen these experimental contexts, CGN is on par or better than Enformer in 5 out of 10 settings.


\subsection{CGN generalises across individuals, tissues, assays, and long-range interactions}
\label{sec:Generalisation}

To systematically evaluate the predictions made by CGN, we employ a more difficult holdout split than previous work \cite{epcot,basenji2,ocelot}. We leverage the ENTEx dataset to assess generalisation across donors, epigenetic contexts, and sequence similarity (rather than simpler chromosomal validation). The holdout framework is compared to existing models in Table~\ref{tab:holdout_properties} and explained in detail in Section~\ref{sec:Materials and methods}.
\begin{table}[h]
  \caption{Properties of the holdout set used to both validate and test existing models.}
  \label{tab:holdout_properties}
  \centering
  \begin{tabular}{ lccccc } 
    \toprule
    \textbf{Model} & \textbf{Chromosome} & \textbf{Sequence similarity} & \textbf{Context} & \textbf{Donor} & \textbf{With variants} \\
    \midrule
    Basenji2 &            & \checkmark &            &            &            \\
    Enformer &            & \checkmark &            &            &            \\
    Epcot    & \checkmark &            &            &            &            \\
    Ocelot   &            &            & \checkmark &            &            \\
    CGN   &            & \checkmark & \checkmark & \checkmark & \checkmark \\
    \bottomrule
  \end{tabular}
\end{table}

\begin{figure}[h]
  \begin{subfigure}[b]{0.5\textwidth}
    \centering
    \includegraphics[width=\textwidth]{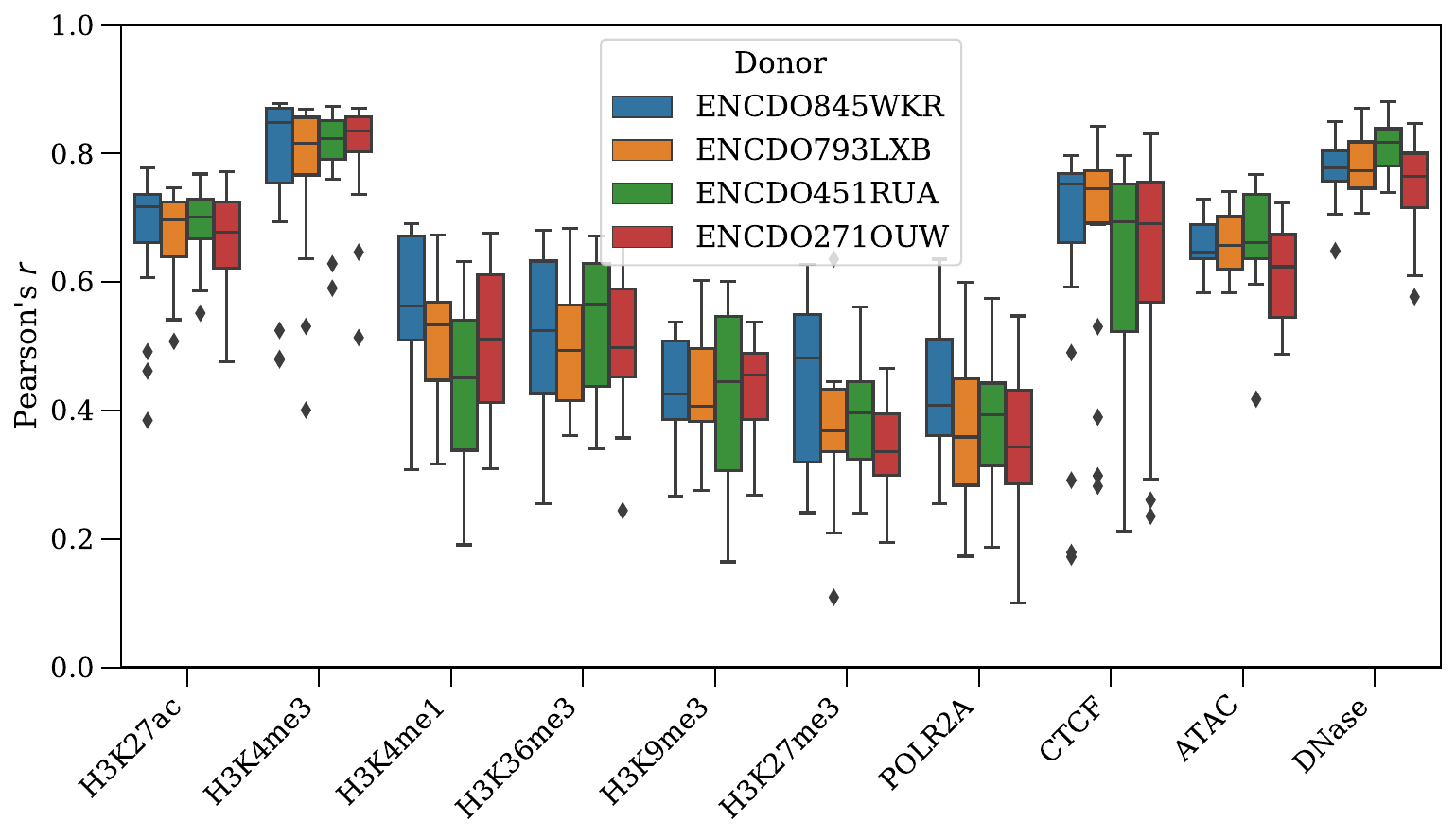}
    \caption{Train on 3 donors, infer on all.}
    \label{fig:cross_tissue_donor_comparison_infer}
  \end{subfigure}%
  \begin{subfigure}[b]{0.5\textwidth}
    \centering
    \includegraphics[width=\textwidth]{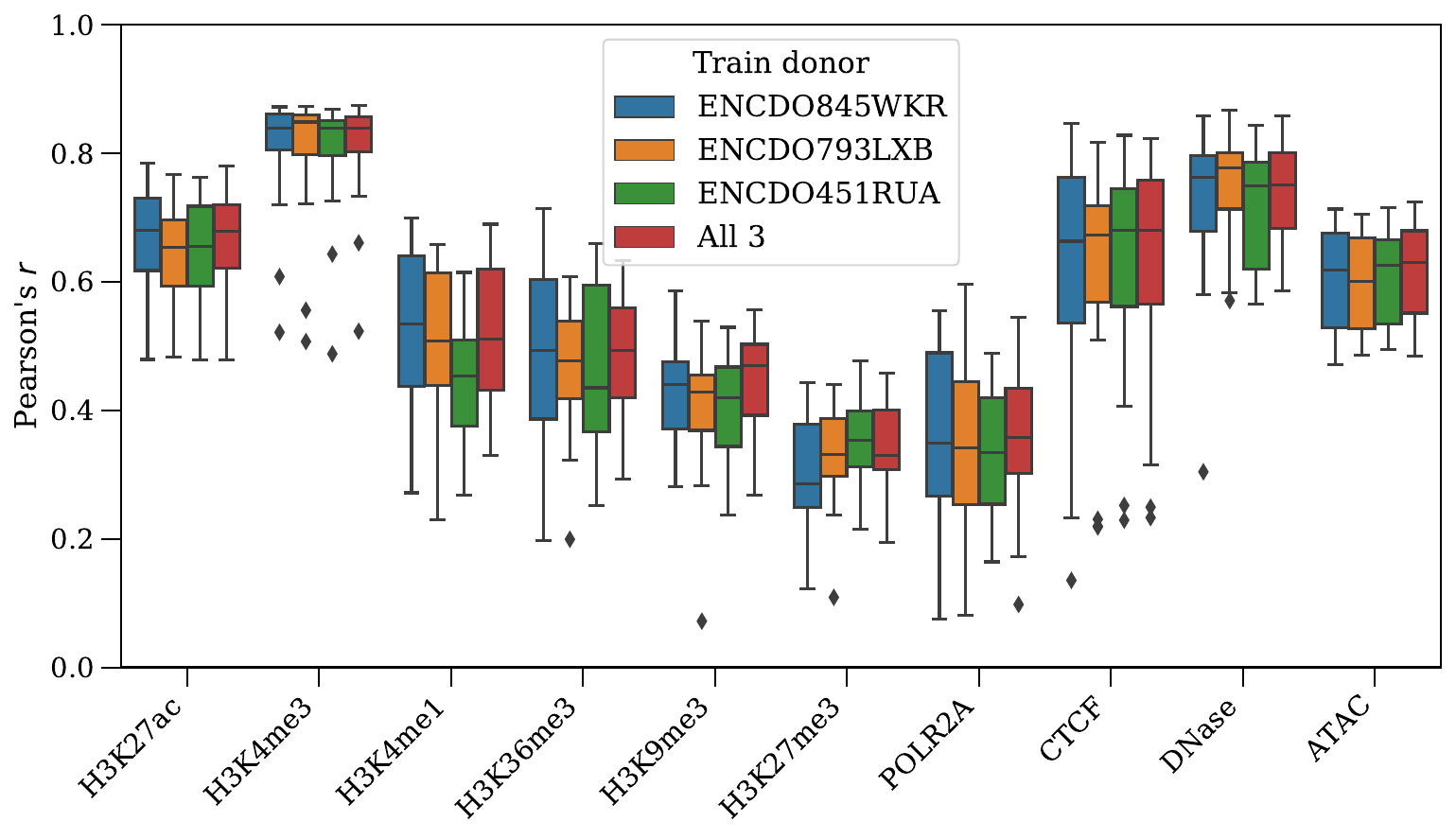}
    \caption{Train on donor subset, infer on held out donor.}
    \label{fig:cross_tissue_donor_comparison_train}
  \end{subfigure}
  \caption{CGN Pearson's $r$ across test set regions, stratified by assay and donor, and averaged across tissues. Outliers were defined as values above or below 1.5 times the interquartile range from the median. Only 2 cross-tissue distributions in (a), H3K27me3 and POL2RA, were found to have a significant difference between donors ($p<0.05$) when using a two-tailed Welch's $t$-test.}
  \label{fig:cross_tissue_donor_comparison}
\end{figure}

We present the first inter-donor epigenetic SM validation in Figure~\ref{fig:cross_tissue_donor_comparison}. When training, we exclude donor ENCDO271OUW before performing test set inference on all 4 donors in Figure~\ref{fig:cross_tissue_donor_comparison_infer}. Similar to prior work, per-donor performance represents generalisation across sequence space, except ENCDO271OUW which demonstrates generalisation across individuals as well. We observe a cross-tissue distribution with comparable performance across donors despite the holdout, with only 2 assays presenting a significant ($p<0.05$) performance drop. We depict the cross-assay stratified results in Figure~\ref{fig:cross_assay_donor_comparison} in the Appendix, where only 1 out of 31 tissues has a significant performance change. In addition, Figure~\ref{fig:cross_tissue_donor_comparison_train} depicts the effect of training on different donor subsets. Training on multiple donors is equal to or better than single-donor training for all assays, although only marginally so. We present extended results in Figure~\ref{fig:train_donor_ablation} in the Appendix, where multi-donor training leads to a notable improvement in the least performant tissues.

\begin{figure}[h]
  \center
  \includegraphics[width=0.6\textwidth]{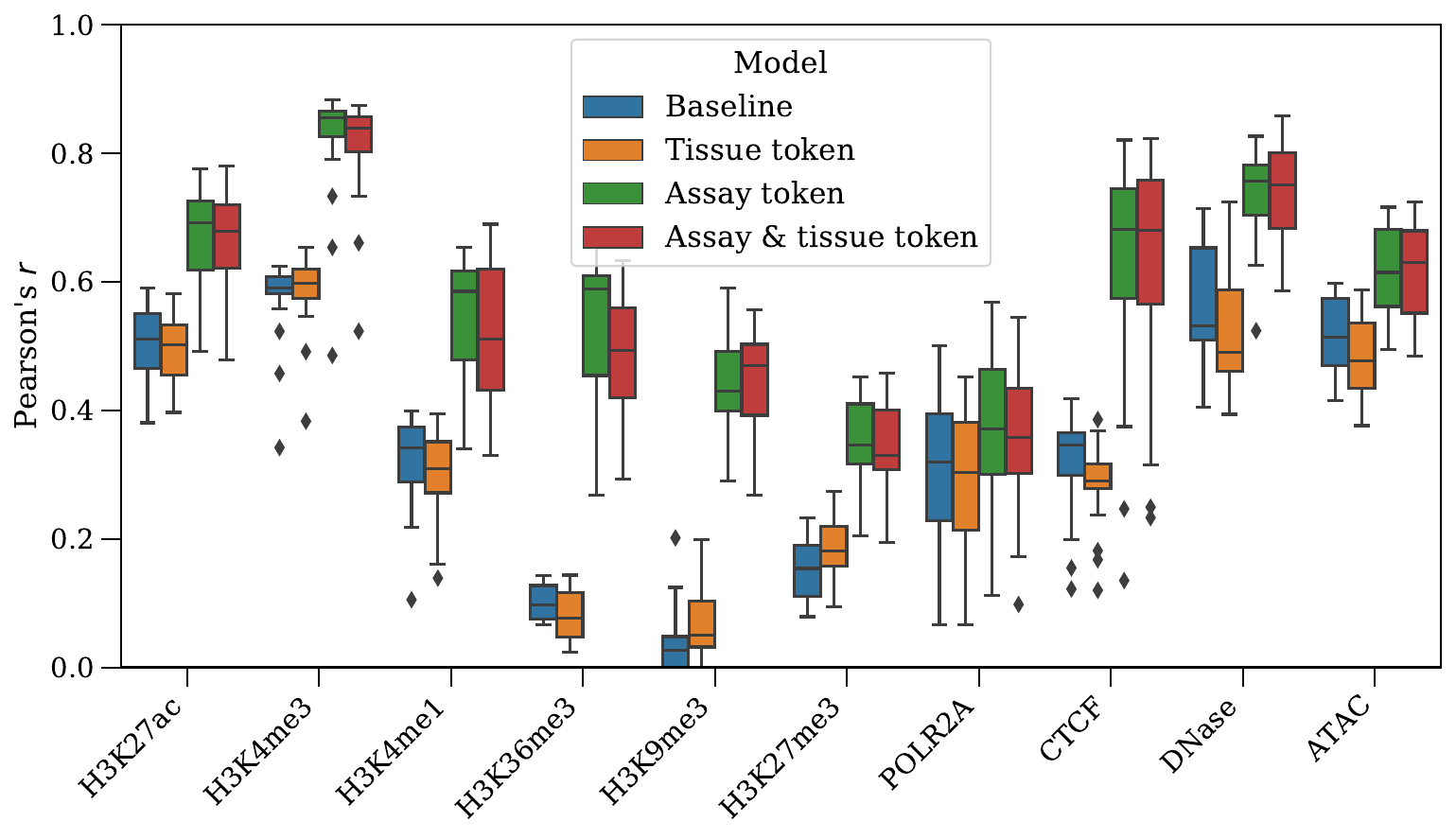}
  \caption{Ablation of the tissue and assay tokens in CGN, stratified by context tokens and assay.}
  \label{fig:token_ablation_cross_tissue}
\end{figure}

\begin{figure}[h]
  \center
  \includegraphics[width=\textwidth]{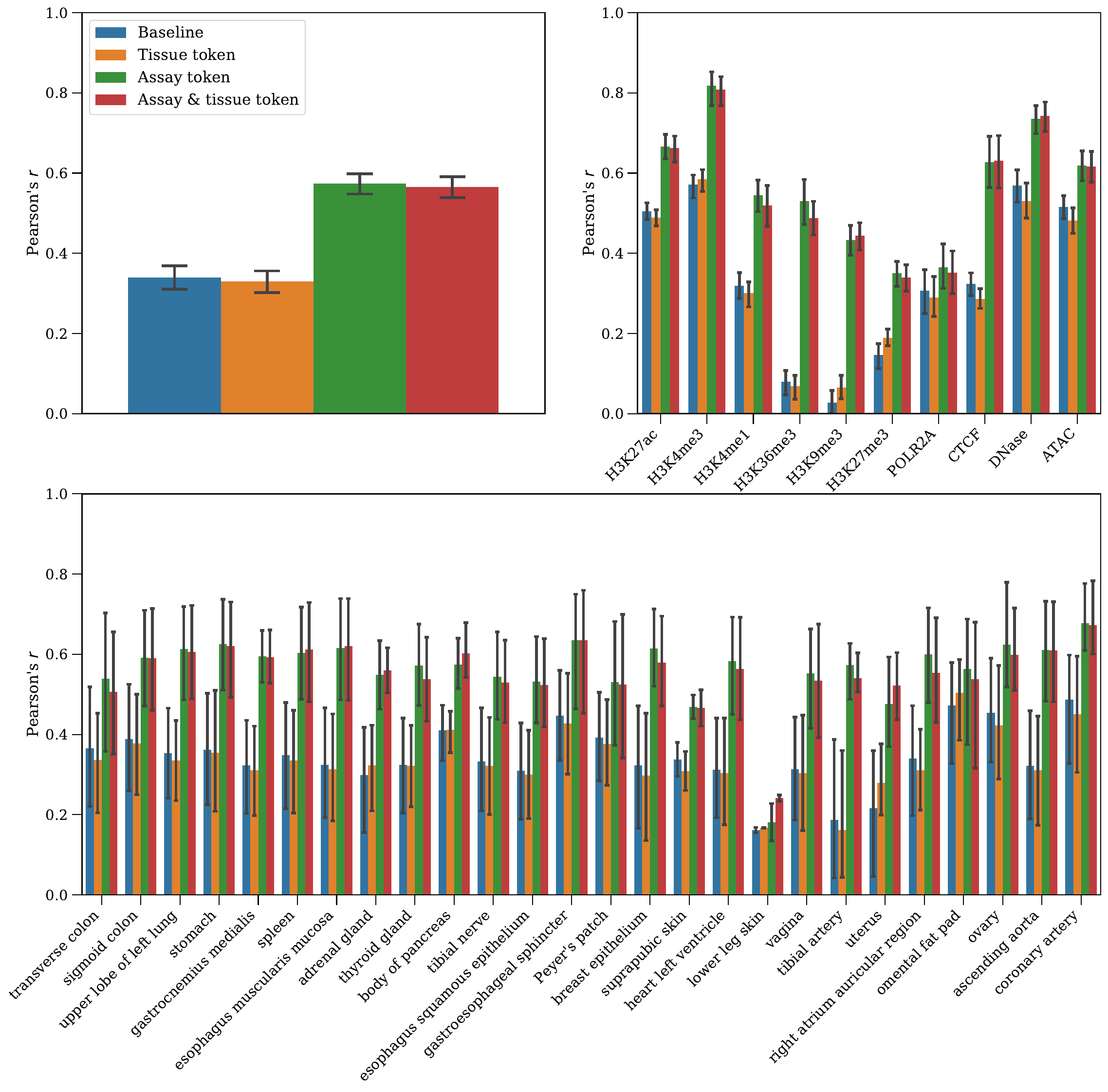}
  \caption{Full ablation of the tissue and assay tokens in CGN, stratified by context token used and tissue or assay. Error bars represent a 95\% confidence interval (CI) around the mean, bootstrapped from 1,000 resamples.}
  \label{fig:token_ablation}
\end{figure}

To isolate the utility of different model and training components contributing to CGN's strong performance, we ablated their effects in Figures~\ref{fig:token_ablation_cross_tissue} and \ref{fig:token_ablation}. We trained models with dummy tissue or assay tokens, or both, to provide the model with no additional information while keeping the architecture fixed (see Section~\ref{sec:Model architecture and training} for details). We observe that the majority of model performance derives from the assay token, which leads to an increase in Pearson's $r$ of approximately 50\%. This is consistent with the aforementioned inter-assay distribution shifts and with prior work opting to use just an assay token \cite{epcot} (see Section~\ref{sec:Discussion}). Additionally, overall performance degrades with the inclusion of a tissue token, regardless of whether an assay token is used. One reason for this could be the strong baseline provided by averaging across tissues which CGN may deviate from for tissue specificity \cite{encodeimputationchallenge}. However, the cross-assay results in Figure~\ref{fig:token_ablation} indicate that the poor performing, and underrepresented, tissues (e.g. \textit{lower leg skin}) may benefit from use of a tissue token.

\subsection{ENTEx short-read SNVs have minimal impact on performance}

\begin{figure}[h]
  \center
  \includegraphics[width=0.6\textwidth]{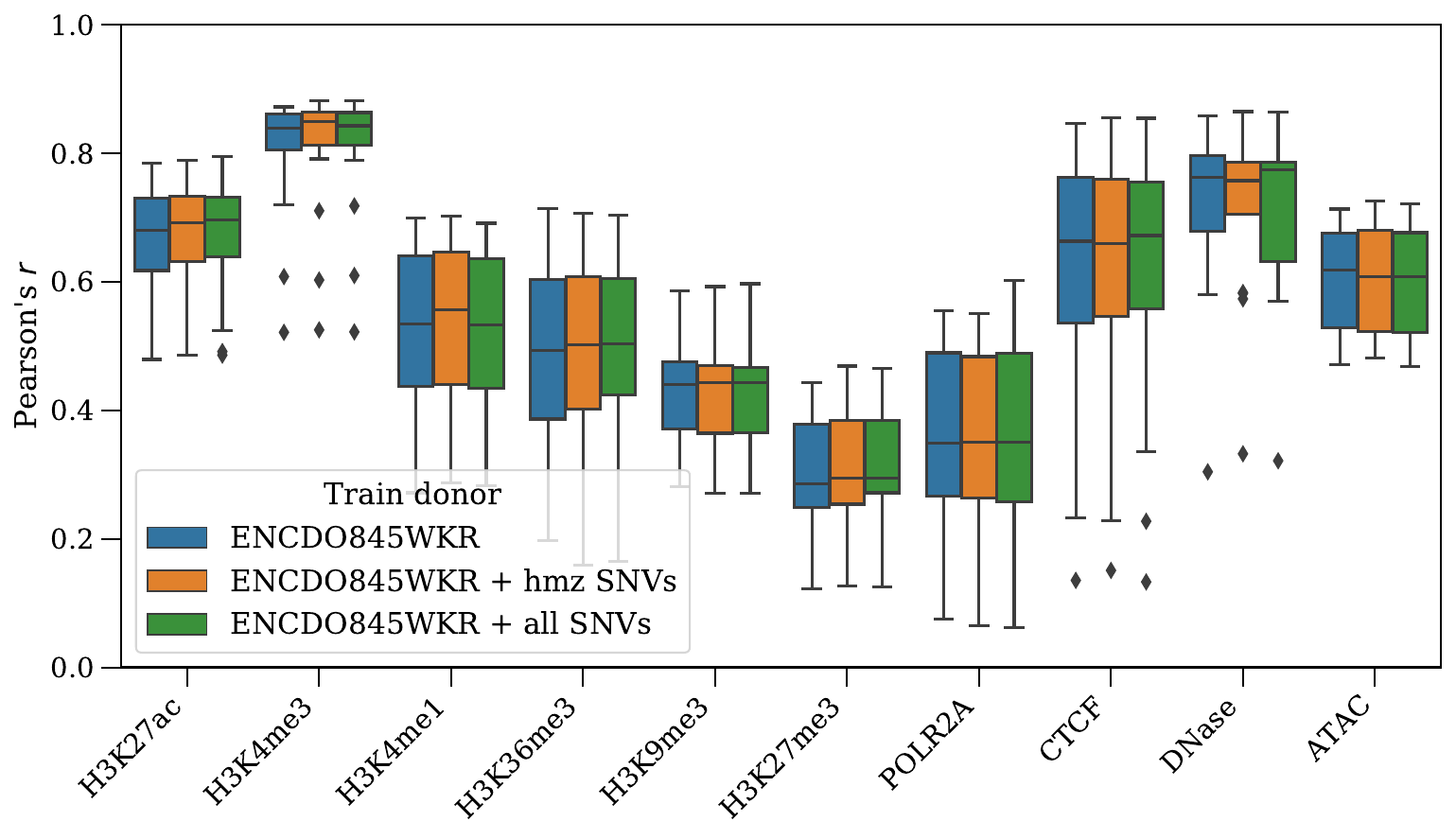}
  \caption{Ablation of SNV usage in CGN, \textit{hmz} denotes homozygosity.}
  \label{fig:snv_ablation}
\end{figure}

In Figure~\ref{fig:snv_ablation}, we present the first results describing the impact of individual variants on epigenetic SMs. We limit our exploration of the short-read variants to single nucleotide variants (SNVs) to avoid ambiguity when mapping DNA sequences with indels to the reference-aligned sequencing assays. We demonstrate in Figure~\ref{fig:snv_ablation} that the use of homozygous and heterozygous SNVs has only a minor impact on performance across assay types in the held out donor. Despite low SNV sensitivity and CGN making a single prediction, rather than an allele-specific (AS) prediction \cite{entex}, the inclusion of heterozygous SNVs is marginally better than homozygous SNVs for a few settings, indicating that some AS activity has been captured. Minimal cross-assay and cross-tissue SNV impact is consistent with recent literature questioning the sensitivity of leading models to variants \cite{sasse2023far,huang2023personal}.

\section{Materials and methods}
\label{sec:Materials and methods}

\subsection{Functional genomics data}
\label{sec:Functional genomics data}
In this work, we studied the effect of four individual genomes across 30 tissues and 10 epigenetic sequencing assays. In particular, we focus on generalising to held out tissue-assay pairs in a held out donor. To delineate such effects, the signal-to-noise ratio in the raw data must be maximised by accounting for known limitations in both the reference genome and each sequencing assay. Therefore, to avoid the effect of spurious alignment errors \cite{pickrell2011false}, or poor reference annotation in the hg38 reference used here \cite{t2t}, we make use of the mappable regions defined during the development of Basenji2 \cite{basenji2}. These regions exclude ENCODE blacklist regions \cite{encodeblacklist}, repeat regions such as telomeres and centromeres \cite{repeatmasker}, and low mappability regions \cite{umapmappability}.

The final dataset consists of 34,021 training, 2,213 validation, and 1,937 test regions from hg38. We use these coordinates to subset both the input DNA sequence and the context-dependent bigWig file from ENTEx which contains the target track. Therefore, during each epoch of training, the model sees the training set regions coupled with a varied set of contexts.

\subsection{Model architecture and training}
\label{sec:Model architecture and training}

The CGN architecture, summarised in Figure~\ref{fig:architecture}, takes as input a sequence of $1536\times128=$196,608 nucleotides which are one-hot encoded, with $\textbf{0}\in\mathbb{R}^4$ representing any base that is not in [A, C, G, T], before using the convolutional and Transformer block architectures from Enformer \cite{enformer}. However, we simplify the Transformer blocks to only use the symmetrical exponential relative positional embeddings. CGN also takes contextual information as an input in the form of 2 special tokens. By default, and in token ablation studies (Figure~\ref{fig:token_ablation}), these tokens are \textit{[NOTISSUE]} and \textit{[NOASSAY]}, optionally replaced by the tissue and assay context, e.g. \textit{stomach} and \textit{CTCF}. These special tokens are embedded in $\mathbb{R}^{64}$, before being projected to the width of the model in question, e.g. $\mathbb{R}^{1,536}$ for the final model. Context embeddings are integrated after the DNA has been down-sampled to a sequence length of 1,536 by the convolutional blocks. We experimented with prepending the context embedding to the DNA embedding, but find that simply adding the tissue, assay, and compressed DNA embeddings together performs better. At the end of the Transformer stack, $1536-640=896$ central bins are cropped from the centre of the sequence for regression. As contextual information is integrated with the model input, rather than output, we use a simple model head consisting of a projection which does not change the representation dimension, a GeLU nonlinearity \cite{gelu}, and a final projection to $\mathbb{R}$ with a softplus activation \cite{softplus}. This stands in contrast to Basenji2 and Enformer where a large intermediate dimension is used before the final projection to several thousand output tracks \cite{basenji2,enformer}. We directly optimise for correlation, using one minus Pearson's $r$ as the loss function because, without a scale-invariant loss function, we find that optimisation is hampered by the contrasting output distributions across assays. As a result, CGN produces a unitless output where absolute scale can be ignored.

For a given sample, we slice 196,608bp from hg38, sample a tissue and assay context pair, slice the context-relevant bigWig in the cropped region, and optionally insert ENTEx SNVs into the DNA sequence. In line with the literature \cite{basenji2}, we use shift augmentation of $\pm3$bp and reverse complement augmentation during training time, although all results reported here are based on a single forward pass at validation and test time (for all models). Targets are dynamically constructed by setting NaN values to zero, averaging over windows of 128bp (simplifying the Gaussian smoothing and soft clipping in prior work \cite{basenji2}), and so no negative values are present.

\begin{wrapfigure}{r}{0.5\textwidth}
  \center
  \vspace{-8mm}
  \includegraphics[width=0.5\textwidth]{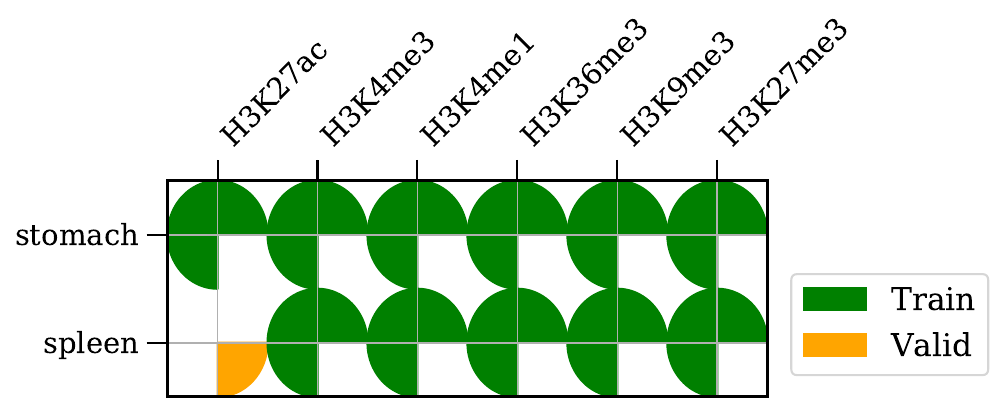}
  \caption{Schematic of the CGN holdout scheme. Circle quadrants represent ENTEx donors.}
  \label{fig:small_grid}
  \vspace{-2mm}
\end{wrapfigure}

We construct the validation and test set by holding out ENTEx donor ENCDO271OUW, a fixed set of tissue-assay pairs (see Figure~\ref{fig:small_grid}), and mappable regions of hg38 split by sequence similarity \cite{basenji2}. When training, donors are provided with their respective variant call format (VCF) file as well as a subset of tissues and assays, e.g. [\textit{stomach}, \textit{spleen}] or [\textit{CTCF}, \textit{H3K27ac}]. At each iteration, we randomly select a tissue-assay pair from the list of training contexts. We weight this selection process by the inverse of the product of the tissue and assay frequencies to encourage stronger performance on underrepresented pairs. After context selection, in contrast to prior work where replicate experiments are merged \cite{basenji2}, when training we randomly choose the bigWig file of a replicate from ENTEx, avoiding selection of any control experiments present in the database. At validation and test time, we evaluate the first replicate for each context. We exclusively used the validation regions during model development and ran inference on the test regions once to generate final results.

CGN was trained for 250,000 steps using a batch size of 64 and the AdamW optimizer with learning rate $10^{-4}$, $\beta_1=0.9$, $\beta_2=0.999$, $\epsilon=10^{-6}$, and gradients clipped to 0.2 on 32 NVIDIA Tesla V100s for 12 days \cite{adamw}. Ablation models had batch size 32, 768 channels, and were trained for 150,000 steps with $\beta_2=0.98$ and gradients clipped to 1.0. Here, channels refers to model width at the end of the convolutional stack and 8 times the value dimension in the multi-head attention component of the Transformer blocks. We varied $\beta_2$ and the gradient clipping between ablation models and the final model as we saw instability in the loss due to the sparse Deoxyribonuclease (DNase) signal.

\subsection{Training and evaluating on individual genomes}
\label{sec:Training and evaluating on individual genomes}
To make use of all individual SNVs in the input sequences, we used the variants called by the ENTEx bioinformatics pipeline \cite{entex}. To avoid technical artefacts, we use variants from the short read sequencing because long reads for different individuals were constructed using different sequencing platforms. As they were available for all donors, from the set of possible short-read calls, we used variants called from the phased Illumina NextSeq 500 reads. During training, after sampling a mappable region, as variants have already been filtered for quality, we integrated the entire set of corresponding SNVs for an individual from their Variant Call Format (VCF) file. We use variant coordinates after applying any shift augmentation and before applying reverse complement augmentation.


\section{Discussion}
\label{sec:Discussion}

\textit{In silico} prediction of the vast combined space of epigenetic assays and cell types is a crucial step toward rapid understanding of regulatory genomics. Due to our reformulation of assay context integration, CGN outperforms the prior state of the art model Enformer across the majority of settings while offering a clear path to expansion further into both tissue and assay space. We also demonstrated that CGN produces competitive predictions in unseen contexts, linking epigenetic sequence models with ENCODE interpolation models \cite{ocelot,basenji2,enformer,avocado}. In contrast to previous approaches, we enhanced long-range sequence embeddings with contextual information in the input space, leading to improved cross-gene correlation which predominantly derived from the assay embedding. We also presented the first use of individual variants from the ENTEx dataset in epigenetic model training.

Although multiple recent models have focussed on epigenetic signals \cite{ebert,ocelot}, the closest approach to ours is that of Epcot \cite{epcot} (see Table~\ref{tab:model_properties}), where multiple epigenetic assay types are predicted by cross-attending to assay embeddings. The authors also augment the fixed nature of the genome across cells with an accessibility signal (DNase-seq/ATAC-seq), which can be interpreted as a continuous cell type representation, while multitask pre-training followed by fine-tuning is used for optimisation. Here we present a simpler approach, replacing the accessibility signal with learnt tissue embeddings specified only by name, a bridging step toward continuous tissue specification via natural language without the need for additional assays. We also use a single multitask training phase which allows zero-shot generalisation to unseen tissue-assay pairs while matching the longest epigenetic sequence model context length to date instead of the 1,600bp regions used in Epcot.


Despite strong correlation, CGN presents multiple future opportunities for research. The discrete cross-tissue and cross-assay generalisation in CGN exposes the model to the tissue definitions in the ENCODE database. In future, a combination of discrete and continuous tissue signals could be considered. Continuous signals could be derived from NLP embeddings of all assay-related information, as well as optional use of any available accessibility signals. For example, using an NLP embedding of donor age, a feature which is readily available in the ENCODE metadata, is likely to have a strong influence on prediction of whole-genome bisulfite sequencing for methylation detection \cite{horvath2013dna}. Moreover, as our results show assay tokens are influential, more information may be gained by carefully embedding the conditions leading to the target signal. In the long term, it may be possible to embed the entire protocol and bioinformatics pipeline used to generate an output bigWig file with a language model. Unlike the epigenetic assays used here, future work could also expand CGN's output space to focus on gene expression prediction as well, rendering CGN even more comparable to Enformer and similar sequence-to-expression models.

Crucially, our use of individual variants highlighted their minimal impact when used during train time. Despite the well known low-mappability regions of hg38, current models either align directly to it or, in the case of ENTEx, align to individual genomes and then liftover back to the reference. In future, our work could be expanded by using the full set of variants for individual genome construction, model inputs, and model outputs. Given current evidence of minimal cross-individual correlation for leading models \cite{sasse2023far,huang2023personal}, we believe this is a promising route to personalised predictions from epigenetic sequence models.






\clearpage

\bibliography{ms}

\begin{thebibliography}{10}

\bibitem{basenji2}
David~R Kelley.
\newblock Cross-species regulatory sequence activity prediction.
\newblock {\em PLoS computational biology}, 16(7):e1008050, 2020.

\bibitem{enformer}
{\v{Z}}iga Avsec, Vikram Agarwal, Daniel Visentin, Joseph~R Ledsam, Agnieszka
  Grabska-Barwinska, Kyle~R Taylor, Yannis Assael, John Jumper, Pushmeet Kohli,
  and David~R Kelley.
\newblock Effective gene expression prediction from sequence by integrating
  long-range interactions.
\newblock {\em Nature methods}, 18(10):1196--1203, 2021.

\bibitem{epcot}
Zhenhao Zhang, Fan Feng, Yiyang Qiu, and Jie Liu.
\newblock A generalizable framework to comprehensively predict epigenome,
  chromatin organization, and transcriptome.
\newblock {\em Nucleic Acids Research}, page gkad436, 2023.

\bibitem{ocelot}
Hongyang Li and Yuanfang Guan.
\newblock Asymmetric predictive relationships across histone modifications.
\newblock {\em Nature machine intelligence}, 4(3):288--299, 2022.

\bibitem{huang2023personal}
Connie Huang, Richard Shuai, Parth Baokar, Ryan Chung, Ruchir Rastogi, Pooja
  Kathail, and Nilah~M Ioannidis.
\newblock Personal transcriptome variation is poorly explained by current
  genomic deep learning models.
\newblock {\em bioRxiv}, pages 2023--06, 2023.

\bibitem{sasse2023far}
Alexander Sasse, Bernard Ng, Anna Spiro, Shinya Tasaki, David~A Bennett,
  Christopher Gaiteri, Philip~L De~Jager, Maria Chikina, and Sara Mostafavi.
\newblock How far are we from personalized gene expression prediction using
  sequence-to-expression deep neural networks?
\newblock {\em bioRxiv}, pages 2023--03, 2023.

\bibitem{lecun1995convolutional}
Yann LeCun, Yoshua Bengio, et~al.
\newblock Convolutional networks for images, speech, and time series.
\newblock {\em The handbook of brain theory and neural networks},
  3361(10):1995, 1995.

\bibitem{vaswani2017attention}
Ashish Vaswani, Noam Shazeer, Niki Parmar, Jakob Uszkoreit, Llion Jones,
  Aidan~N Gomez, {\L}ukasz Kaiser, and Illia Polosukhin.
\newblock Attention is all you need.
\newblock {\em Advances in neural information processing systems}, 30, 2017.

\bibitem{seq2cells}
Ron Schwessinger, Jacob Deasy, Rob~T Woodruff, Stephen Young, and Kim~M
  Branson.
\newblock Single-cell gene expression prediction from dna sequence at large
  contexts.
\newblock {\em bioRxiv}, pages 2023--07, 2023.

\bibitem{entex}
Joel Rozowsky, Jiahao Gao, Beatrice Borsari, Yucheng~T Yang, Timur Galeev,
  Gamze G{\"u}rsoy, Charles~B Epstein, Kun Xiong, Jinrui Xu, Tianxiao Li,
  et~al.
\newblock The en-tex resource of multi-tissue personal epigenomes \&
  variant-impact models.
\newblock {\em Cell}, 186(7):1493--1511, 2023.

\bibitem{chainofthought}
Jason Wei, Xuezhi Wang, Dale Schuurmans, Maarten Bosma, Fei Xia, Ed~Chi, Quoc~V
  Le, Denny Zhou, et~al.
\newblock Chain-of-thought prompting elicits reasoning in large language
  models.
\newblock {\em Advances in Neural Information Processing Systems},
  35:24824--24837, 2022.

\bibitem{whisper}
Alec Radford, Jong~Wook Kim, Tao Xu, Greg Brockman, Christine McLeavey, and
  Ilya Sutskever.
\newblock Robust speech recognition via large-scale weak supervision.
\newblock In {\em International Conference on Machine Learning}, pages
  28492--28518. PMLR, 2023.

\bibitem{multimodal}
S{\"o}ren~Richard Stahlschmidt, Benjamin Ulfenborg, and Jane Synnergren.
\newblock Multimodal deep learning for biomedical data fusion: a review.
\newblock {\em Briefings in Bioinformatics}, 23(2):bbab569, 2022.

\bibitem{avocado}
Jacob Schreiber, Timothy Durham, Jeffrey Bilmes, and William~Stafford Noble.
\newblock Avocado: a multi-scale deep tensor factorization method learns a
  latent representation of the human epigenome.
\newblock {\em Genome biology}, 21(1):1--18, 2020.

\bibitem{eDICE}
Alex Hawkins-Hooker, Giovanni Vison{\`a}, Tanmayee Narendra, Mateo
  Rojas-Carulla, Bernhard Sch{\"o}lkopf, and Gabriele Schweikert.
\newblock Getting personal with epigenetics: towards individual-specific
  epigenomic imputation with machine learning.
\newblock {\em Nature Communications}, 14(1):4750, 2023.

\bibitem{encodeimputationchallenge}
Jacob Schreiber, Carles Boix, Hongyang Li, Yuanfang Guan, Chun-Chieh Chang,
  Jen-Chien Chang, Alex Hawkins-Hooker, Bernhard Sch{\"o}lkopf, Gabriele
  Schweikert, Mateo~Rojas Carulla, et~al.
\newblock The encode imputation challenge: a critical assessment of methods for
  cross-cell type imputation of epigenomic profiles.
\newblock {\em Genome Biology}, 24(1):1--22, 2023.

\bibitem{pickrell2011false}
Joseph~K Pickrell, Daniel~J Gaffney, Yoav Gilad, and Jonathan~K Pritchard.
\newblock False positive peaks in chip-seq and other sequencing-based
  functional assays caused by unannotated high copy number regions.
\newblock {\em Bioinformatics}, 27(15):2144--2146, 2011.

\bibitem{t2t}
Sergey Nurk, Sergey Koren, Arang Rhie, Mikko Rautiainen, Andrey~V Bzikadze,
  Alla Mikheenko, Mitchell~R Vollger, Nicolas Altemose, Lev Uralsky, Ariel
  Gershman, et~al.
\newblock The complete sequence of a human genome.
\newblock {\em Science}, 376(6588):44--53, 2022.

\bibitem{encodeblacklist}
Haley~M Amemiya, Anshul Kundaje, and Alan~P Boyle.
\newblock The encode blacklist: identification of problematic regions of the
  genome.
\newblock {\em Scientific reports}, 9(1):9354, 2019.

\bibitem{repeatmasker}
Darryl Nishimura.
\newblock Repeatmasker.
\newblock {\em Biotech Software \& Internet Report}, 1(1-2):36--39, 2000.

\bibitem{umapmappability}
Mehran Karimzadeh, Carl Ernst, Anshul Kundaje, and Michael~M Hoffman.
\newblock Umap and bismap: quantifying genome and methylome mappability.
\newblock {\em Nucleic acids research}, 46(20):e120--e120, 2018.

\bibitem{gelu}
Dan Hendrycks and Kevin Gimpel.
\newblock Gaussian error linear units (gelus).
\newblock {\em arXiv preprint arXiv:1606.08415}, 2016.

\bibitem{softplus}
Charles Dugas, Yoshua Bengio, Fran{\c{c}}ois B{\'e}lisle, Claude Nadeau, and
  Ren{\'e} Garcia.
\newblock Incorporating second-order functional knowledge for better option
  pricing.
\newblock {\em Advances in neural information processing systems}, 13, 2000.

\bibitem{adamw}
Ilya Loshchilov and Frank Hutter.
\newblock Decoupled weight decay regularization.
\newblock {\em arXiv preprint arXiv:1711.05101}, 2017.

\bibitem{ebert}
Meredith~V Trotter, Cuong~Q Nguyen, Stephen Young, Rob~T Woodruff, and Kim~M
  Branson.
\newblock Epigenomic language models powered by cerebras.
\newblock {\em arXiv preprint arXiv:2112.07571}, 2021.

\bibitem{horvath2013dna}
Steve Horvath.
\newblock Dna methylation age of human tissues and cell types.
\newblock {\em Genome biology}, 14(10):1--20, 2013.

\end{thebibliography}
\bibliographystyle{unsrt}

\clearpage

\section{Appendix}


\subsection{Extended prediction figures}

\begin{figure}[h]
  \center
  \includegraphics[width=\textwidth]{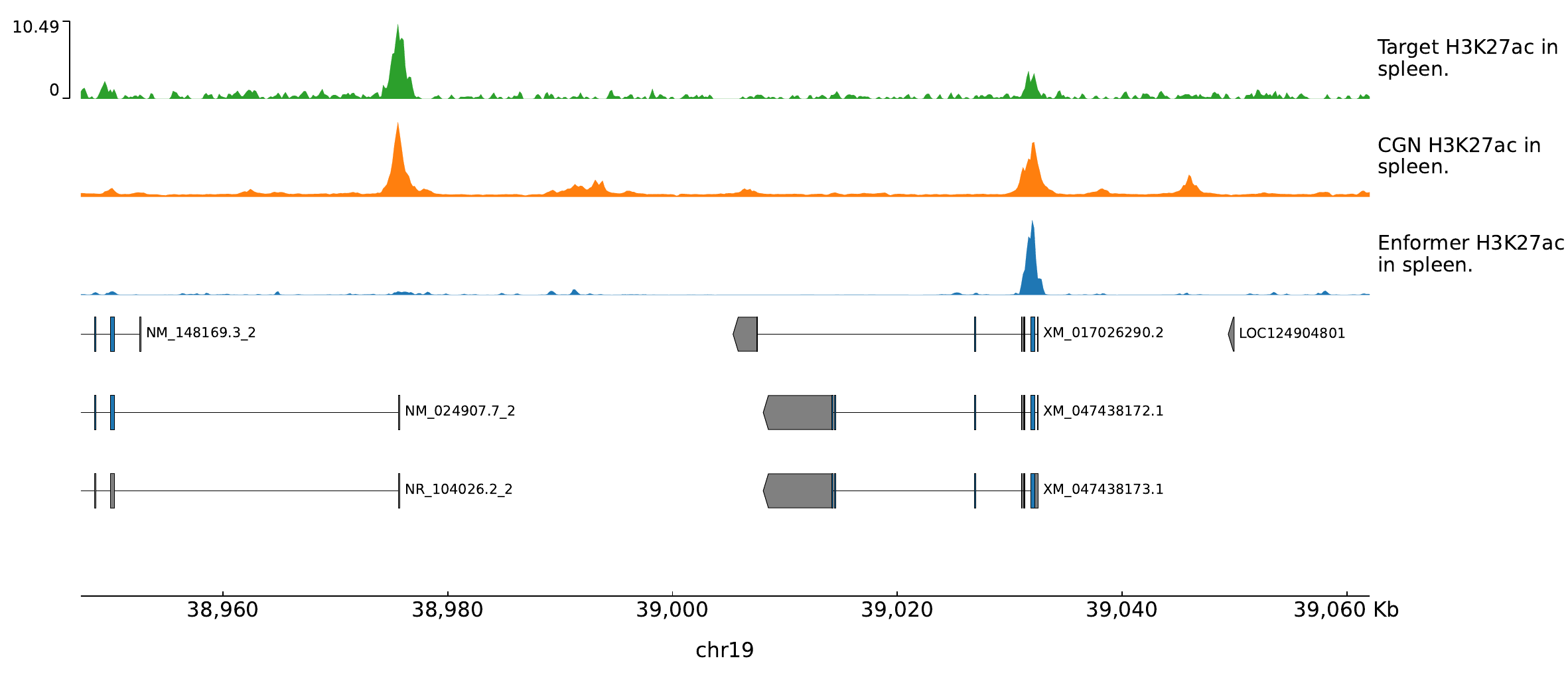}
  \caption{An example test set region where CGN captures an increase in log fold change over control (LFOC) at the transcription start site of FBXO17 and Enformer does not. Enformer local Pearson's $r=0.476$ and CGN local Pearson's $r=0.794$.}
  \label{fig:enformer_miss}
\end{figure}

\begin{figure}[h]
  \label{fig:heldout_donor_correlations}
  \center
  \includegraphics[width=\textwidth]{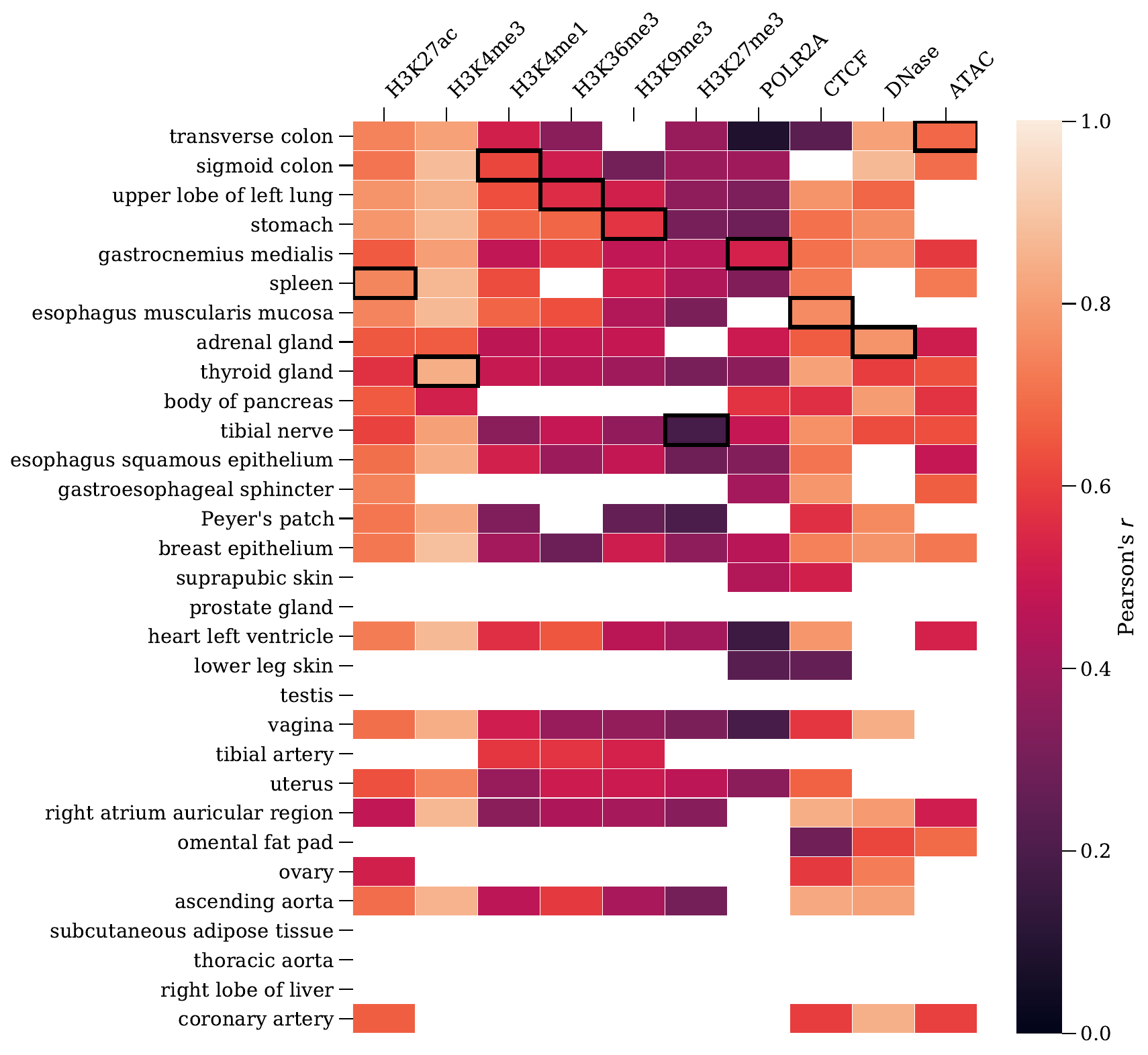}
  \caption{Test set Pearson's $r$ for held out ENTEx donor ENCDO271OUW used to assess CGN. Missing and black-border cells indicate missing and held out tissue-assay pairs respectively. None of these pairs are used in training. The ENTEx dataset used includes 6 histone marks, 2 transcription factors, as well as DNase-seq and ATAC-seq, all across 31 tissues. The resultant model can predict in 310 contexts, 186 of which are used for performance assessment, while the remaining 124 can be imputed.}
\end{figure}

\clearpage

\subsection{Extended ablations}

\begin{figure}[h]
  \center
  \includegraphics[width=\textwidth]{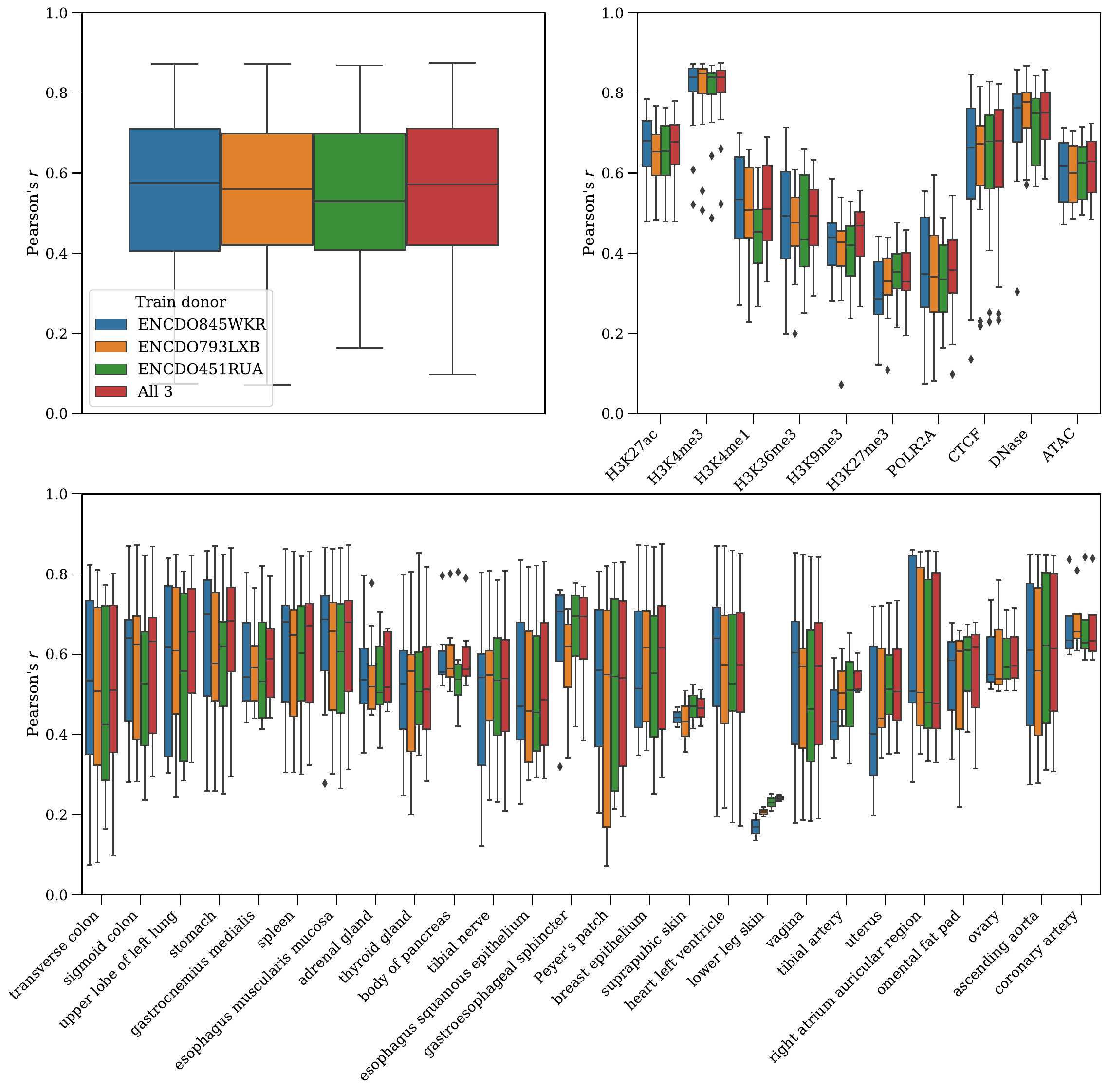}
  \caption{CGN Pearson's $r$ across test set regions, stratified by donor and tissue or assay. Outliers were defined as values above or below 1.5 times the interquartile range from the median.}
  \label{fig:train_donor_ablation}
\end{figure}

\begin{figure}[h]
  \center
  \includegraphics[width=\textwidth]{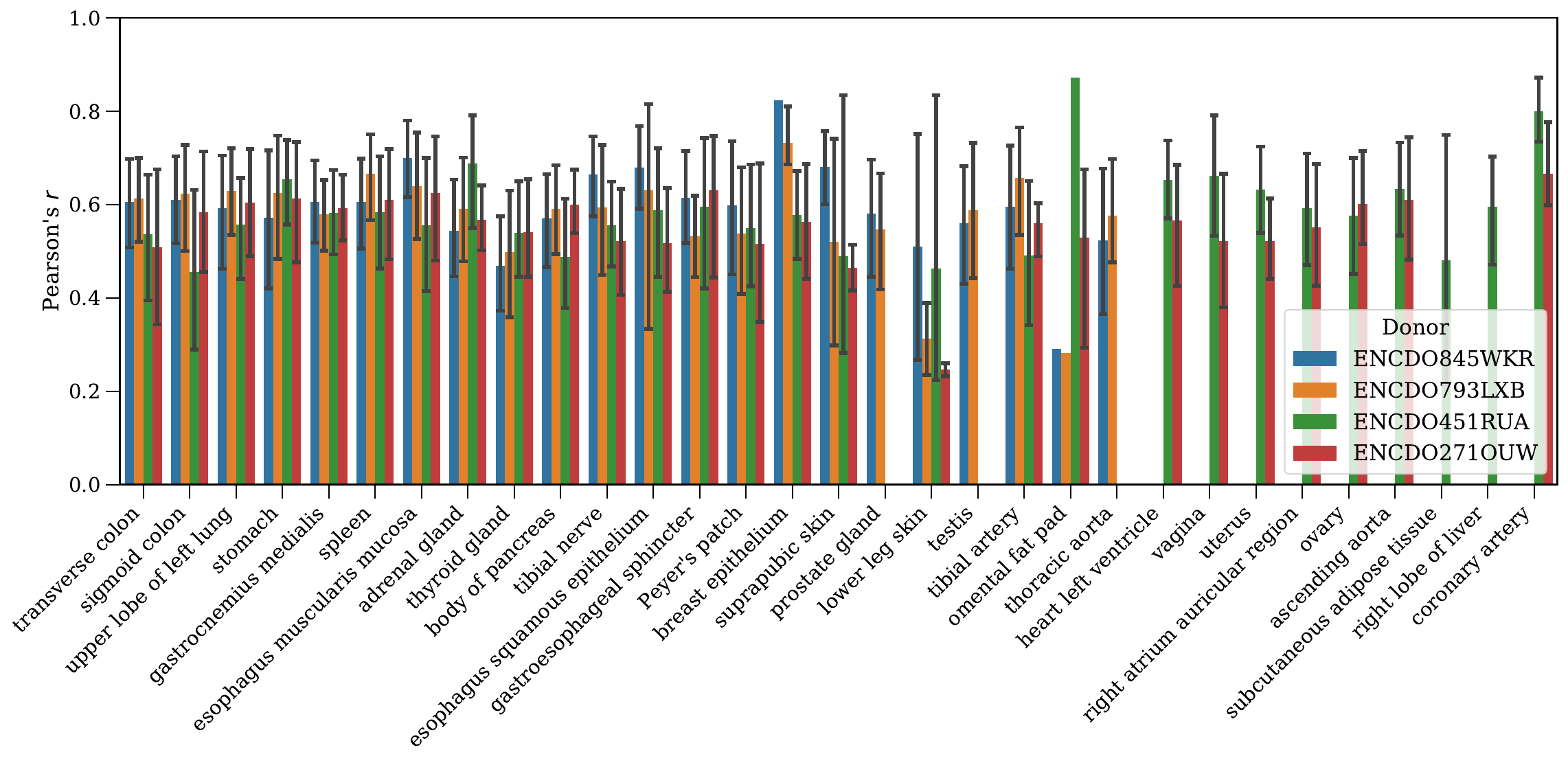}
  \caption{CGN Pearson's $r$ across test set regions, stratified by tissue and inference donor, and averaged across assays. Similar to prior work, per-donor performance represents generalisation across sequence space, except ENCDO271OUW which demonstrates generalisation across individuals as well. Error bars represent a 95\% confidence interval (CI) around the mean, bootstrapped from 1,000 resamples. Donor cross-assay distribution shifts were compared with a two-tailed Welch's $t$-test and all found to be insignificant ($p>0.05$).}
  \label{fig:cross_assay_donor_comparison}
\end{figure}

\begin{figure}[h]
  \center
  \includegraphics[width=\textwidth]{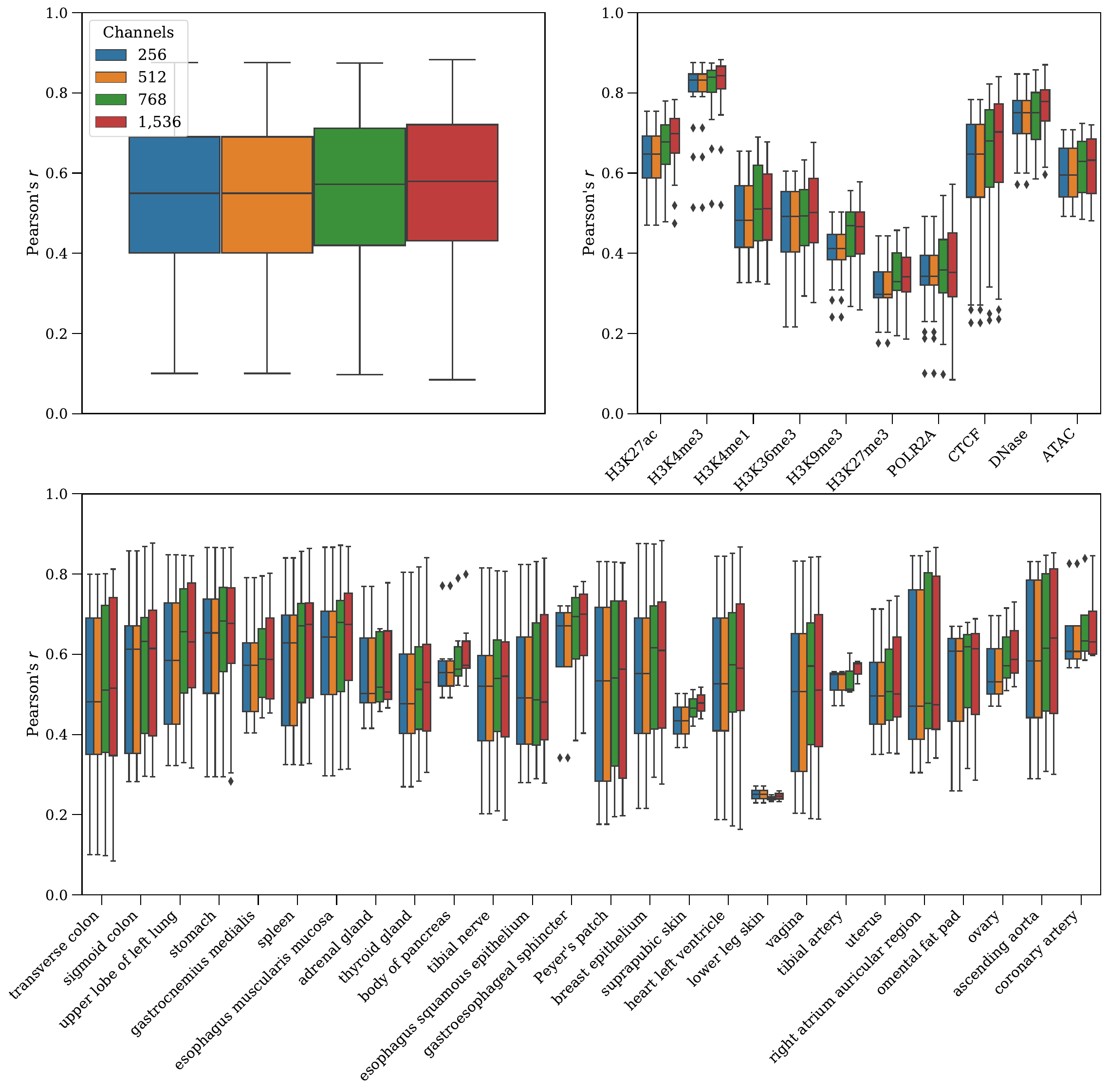}
  \caption{Ablation of the number of channels used in CGN. As in \cite{enformer}, channels refers to the number of channels at the end of the convolutional stack and 8 times the subsequent value dimension in the multi-head attention component of the Transformer blocks. The increasing performance from 256 to 768 is consistent with that reported in Enformer and justified the large final model size in red.}
  \label{fig:channels_ablation}
\end{figure}

\end{document}